\documentclass[sigconf]{acmart}

 \renewcommand\footnotetextcopyrightpermission[1]{}

\usepackage{gensymb}
\usepackage{mathtools}
\usepackage{subcaption}
\usepackage{graphicx} 
\usepackage[skip=3pt]{caption} 

\usepackage{ifthen}
\usepackage{xcolor}
\definecolor{myhl}{rgb}{1,0.5,0}
\newboolean{showblocks}
\setboolean{showblocks}{false} 
\newcommand{\colorblock}[1]{\ifthenelse{\boolean{showblocks}}{\textcolor{myhl}{#1}}{#1}}

\setcopyright{acmcopyright}
\copyrightyear{2024}
\acmYear{2024}
\acmDOI{XXXXXXX.XXXXXXX}

\acmConference[IPSN]{The ACM/IEEE International Conference on Information Processing in Sensor Networks}{May 13-16, 2024}{Hong Kong, China}

\acmPrice{15.00}
\acmISBN{978-1-4503-XXXX-X/18/06}

\acmSubmissionID{25}

\AtBeginDocument{%
  \providecommand\BibTeX{{%
    \normalfont B\kern-0.5em{\scshape i\kern-0.25em b}\kern-0.8em\TeX}}}

\begin{document}

\title{Beyond-Voice: Towards Continuous 3D Hand Pose Tracking on Commercial Home Assistant Devices}



\author{Yin Li, Rohan Reddy, Cheng Zhang, Rajalakshmi Nandakumar}
\affiliation{%
  \institution{Cornell University}
  \country{}
}
\email{{yl3243, rr784, chengzhang, rajalakshmi.nandakumar}@cornell.edu}





\begin{abstract}
The surging popularity of home assistants and their voice user interface (VUI) have made them an ideal central control hub for smart home devices.
However, current form factors heavily rely on VUI, which poses accessibility and usability issues; some latest ones are equipped with additional cameras and displays, which are costly and raise privacy concerns. 
These concerns jointly motivate Beyond-Voice, a novel high-fidelity acoustic sensing system that allows commodity home assistant devices to track and reconstruct hand poses continuously. It transforms the home assistant into an active sonar system using its existing onboard microphones and speakers. We feed a high-resolution range profile to the deep learning model that can analyze the motions of multiple body parts and predict the 3D positions of 21 finger joints, bringing the granularity for acoustic hand tracking to the next level. It operates across different environments and users without the need for personalized training data. A user study with 11 participants in 3 different environments shows that Beyond-Voice can track joints with an average mean absolute error of 16.47mm without any training data provided by the testing subject.

\end{abstract}







\maketitle

\section{Introduction}

Commercial home assistant devices, such as Amazon Echo, Google Home, Apple HomePod and Meta Portal, primarily employ voice-user interfaces (VUI) to facilitate verbal speech-based interaction. While the VUIs are generally well received, relying primarily on a speech interface raises (1) accessibility concerns by precluding those with speech disabilities from interacting with these devices and (2) usability concerns stemming from a general misinterpretation of user input due to factors such as non-native speech or background noise ~\cite{pyae2018investigating, masina2020investigating,pyae2019investigating, garg2021learn}. While some of the latest home assistant devices have cameras for motion tracking and displays with touch interfaces, these systems are relatively expensive, not immediately available to millions of existing devices, and also raise privacy concerns.

In this paper, we propose a beyond-voice method of interaction with these devices as a complementary technique to alleviate the accessibility and usability issues of VUI. Our system leverages the existing acoustic sensors of commercial home assistant devices to enable continuous fine-grained hand tracking of a subject. In comparison, current acoustic hand tracking systems ~\cite{li2020fm, mao2019rnn, nandakumar2016fingerio, wang2016device} have insufficient detection granularity, i.e. discrete gestures classification, or localize a single nearest point, or up to 2 points per hand. Our system enables fine-grained multi-target tracking of the hand pose by 3D localizing the 21 individual joints of the hand. This significantly improves the expressiveness of the gestures and user experience of smart speakers in various interaction scenarios, such as i) using continuous gesture commands, like zoom-in or turning a knob to adjust the volume to a specific level. ii) sign language communication without pre-defining gestures in training.

Our system increases the level of detection granularity of acoustic sensing to enable articulated hand pose tracking of the subject by leveraging the existing speaker and microphones in the device. The key idea is to transform the device into an active sonar system. We play inaudible ultrasound chirps (Frequency Modulated Continuous Wave, FMCW) using a speaker and record the reflections using a co-located circular microphone array. By analyzing the time-of-flight in the signal reflected from the moving hand, we can 3D localize the 21 finger joints of the hand.

Building a continuous hand tracking system poses several challenges. First, the system needs to locate the joints in the ambient environment, even in unseen environments. Therefore, we design a signal processing pipeline that can eliminate unwanted reflections and then combine multiple microphones to localize the hand in 3D. Nevertheless, the reflections from joints are entangled making it intractable to separate them with rule-based algorithms, especially in the presence of multi-path noise from moving fingers. Hence, we use a Convolutional Neural Network (CNN) + Long Short-Term Memory (LSTM) model to learn the patterns in the signal reflection of multi-parts, i.e. 3D position of 21 joints. In training, we use a Leap Motion depth camera as ground truth and a curriculum learning (CL) technique to hierarchically pre-train the model.

Secondly, it should work across different ranges and orientations. However, this would require a huge data collection effort to train a system that detects fine-grained absolute positions in a large search space around the device. Therefore, we design a customized data augmentation method for range profile to alleviate this effort, which also helps with overfitting reduction. We prove that training strategies of CL and data augmentation can effectively improve performance with standard pose estimation models.

Finally, the system should work for unseen users without adaptive training using their data, i.e. the model should be generalizable across users, namely user-independent. Since the ground truth cameras in our experiment are usually not available in cheap home assistants, it is not practical to expect individual users to train the system separately. To address this, a one-time extensive training dataset was collected from multiple subjects, which can then directly apply to unseen subjects after deployment. However, it is known that adaptive training with personalized data usually could improve performance. Thereby, we also show the user-adaptive results in case cameras are available in some form-factors.

To evaluate our system, we deploy Beyond-Voice on a development board with similar hardware settings with Amazon Echo Dot 2, one of the most popular smart home assistant devices. Using this prototype, we conducted a user study with 11 users, providing a total of 64 minutes of data, which are carefully selected hand motions that expressively cover the movements of all finger joints. Before the user study, we pre-trained a model using curriculum learning with 40 minutes of data from two researchers excluded from the user study. The system average MAE is 16.47mm (median 14.57mm) in the user-independent study. Hence the system can perform hand tracking without any training data from a new user. If adding two minutes of data from a new user for adaptive training, the MAE can further decrease to 10.36mm (median 9.72mm) in a user-adaptive evaluation. Ultimately, we showcase the system's capability to reconstruct common hand poses, including zooming in with the thumb and index finger and executing sign language gestures as seamlessly supported downstream applications.

In a comprehensive overview, this paper presents the following key contributions:

\begin{itemize}
    \item We develop a novel fine-grained 3D hand tracking system, leveraging the existing acoustic sensors in the home assistant devices. Our continuous tracking of 21 finger joints is unbounded to predefined gestures, enabling versatile downstream applications.
    \item Our deep learning model can generalize across both environments and users without personalized adaptive training.
    \item We evaluated our system with a user study with 11 users in three different environments across different days. The system yields an average MAE of 16.47mm (median 14.57) user-independently.
\end{itemize}



\section{Related Work}


\begin{figure*}
    \begin{minipage}{0.72\textwidth}
         \centering
          \includegraphics[width=0.95\linewidth]{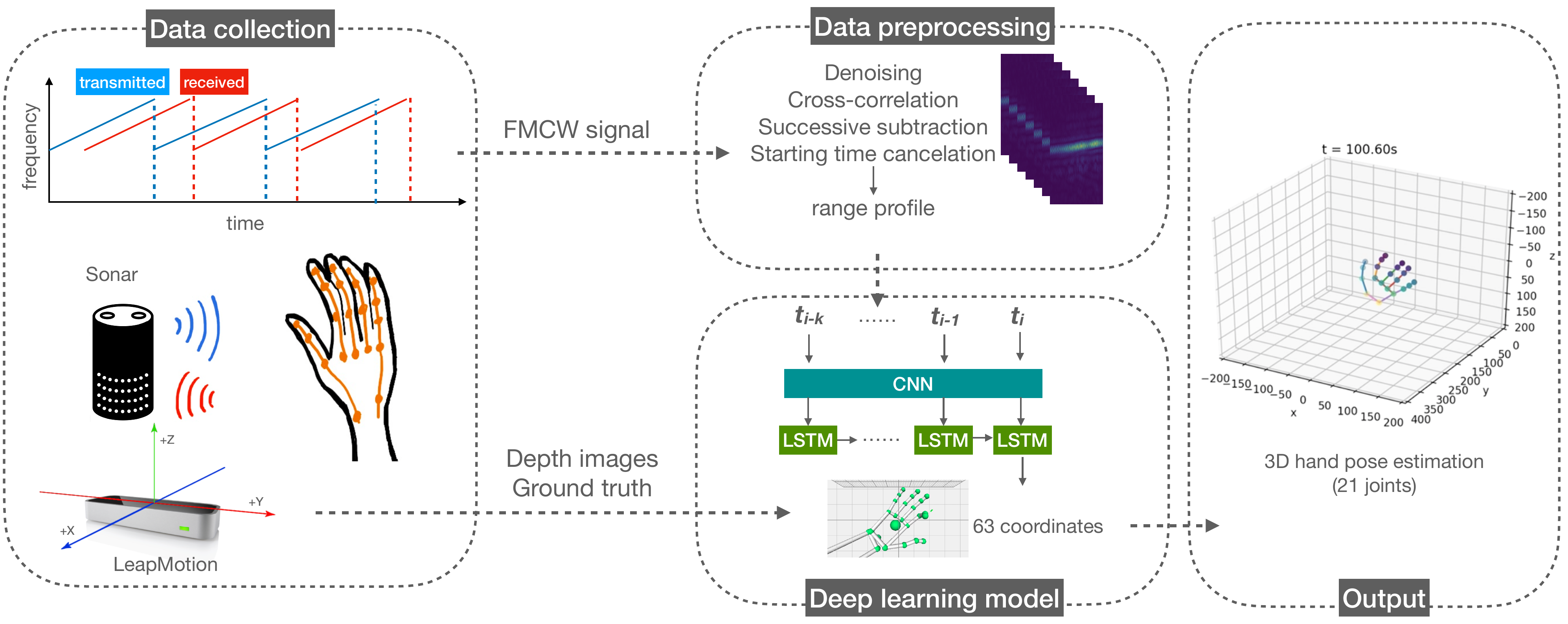}
          \caption{System overview}
          \label{fig:workflow}
    \end{minipage}
    \begin{minipage}{0.27\textwidth}
        \centering
        \vspace{10mm}
        \includegraphics[width=0.85\linewidth]{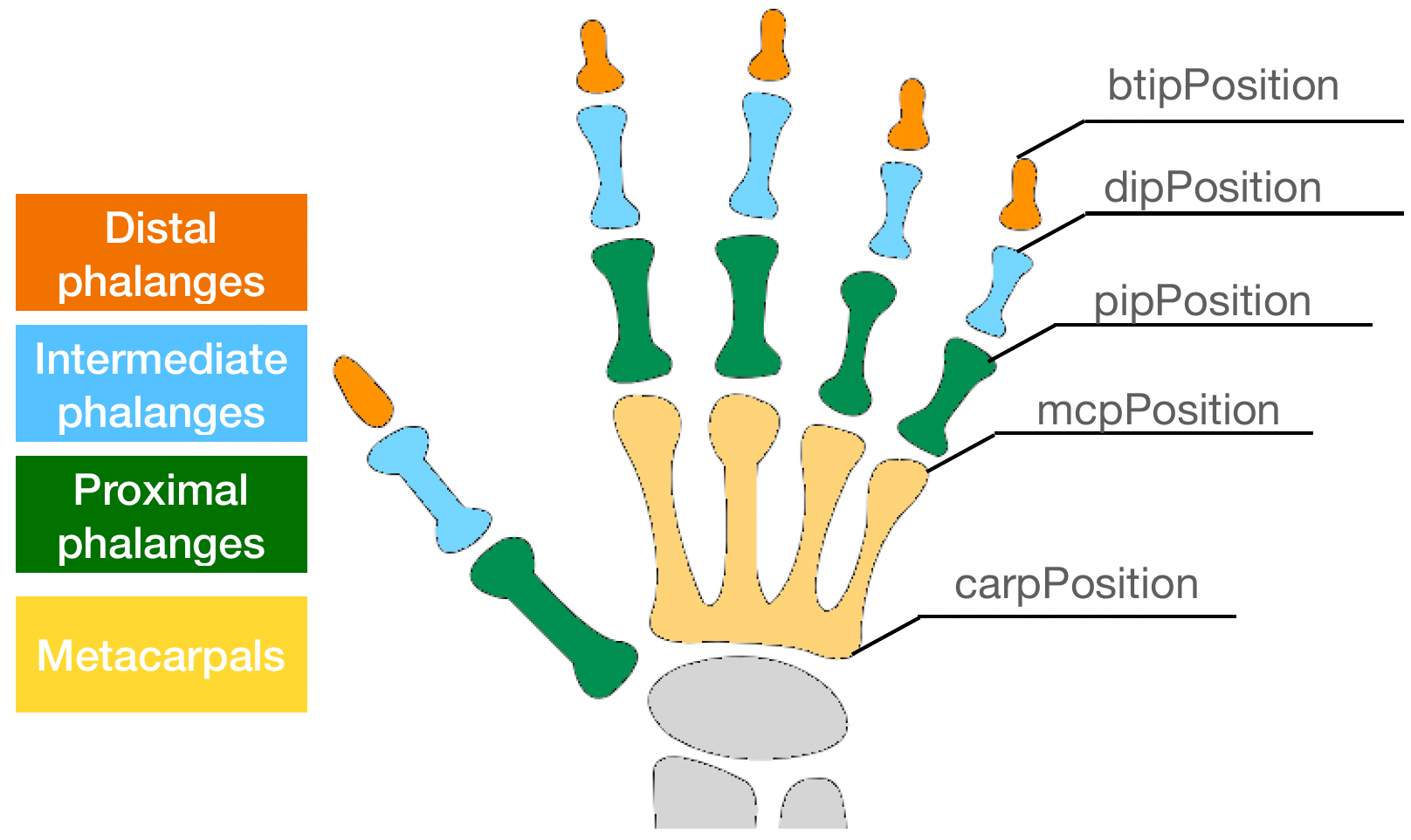}
        \vspace{3mm}
        \caption{We reconstruct 21 joints, mirroring ground truth camera's API. \cite{leapJoints}.}
        \label{fig:skeleton}
    \end{minipage}
\end{figure*}

The existing literature in this field of hand tracking can be categorized into four different types of solutions at a high level: a) Wearable sensor-based solutions; b) Vision-based solutions; c) RF-based solutions; and d) Acoustic solutions.

\subsection{Wearable sensor-based solution} 

To implement high-fidelity hand tracking, wearables are considered one of the most promising solutions. Rings and wristbands, like AuraRing~\cite{parizi2019auraring} and SoundTrack~\cite{zhang2017soundtrak}, can continuously track hand and finger motion using magnetic or acoustic sensors. But only the index finger with the ring on is tracked. FingerTrack\cite{hu2020fingertrak} tracks multiple key points using a wristband with miniature thermal cameras. EMG wristbands~\cite{quivira2018translating, sosin2018continuous, metawrist, liu2021wr} can continuously track pre-defined gesture sequences. Apple added gesture control in their latest Apple Watch, but it only uses IMU to recognize two simple gestures, i.e. pinch and clench. Meta\cite{metaglove} and CyberGlove\cite{cyberglove} build gloves for articulated tracking and haptics. Nevertheless, in general, the wearables are cumbersome and user needs to wear them on and off in between other daily activities.

\subsection{Vision-based solution} 

Hand pose estimation is a well-explored topic in the computer vision community. RGB cameras\cite{simon2017hand, ge20193d, zhang2020mediapipe, Li2022intaghand}, and depth cameras\cite{oikonomidis2011efficient, ge2016robust, wan2019self} can track hand in 3D or 2.5D with under-centimeter error. This development benefits from deep learning and large open-source datasets. They can also generate pseudo data easily using simulation environments such as Unity. In the industry, some mature products such as Kinect\cite{oikonomidis2011efficient}, Oculus Quest\cite{han2020megatrack}, and Leap Motion\cite{leapmotion} are equipped with depth cameras for real-time tracking.

However, since hand gestures are highly self-occluded, vision-based cameras inevitably have limitations on non-line-of-sight parts. Besides, the vision sensors are sensitive to the variability of image-related factors such as illumination, background, occlusion, image resolution, hand orientation, and visual hand characteristics that negatively affect optimal image quality and reduce recognition performance. To handle the problem of occlusion, these works usually need more than one sensor to detect from different angles. For example, \cite{simon2017hand} requires multiple RGB cameras capturing different perspectives of the same scene to improve performance. MEgATrack\cite{han2020megatrack} uses four fisheye monochrome cameras located on the Oculus headset. Nevertheless, in real-world highly-occluded tasks such as hand tracking, performance degrades with limited deployment location and the line-of-sight. Therefore, latest works~\cite{yang2022posekernellifter, yang2022camera} use a fusion of camera and acoustic sensing for body pose estimation.

However, cameras raise privacy concerns, particularly in home-use scenarios. In contrast, our acoustic-sensing-only tracking is a promising alternative, offering comparable accuracy and fidelity in comparison with the existing camera-based systems. While wireless perception has fewer privacy issues.

\subsection{RF-based solution} 

Recently, wireless perception using RF signals is a popular research field. Project Soli\cite{hayashi2021radarnet, wang2016interacting} by Google uses a custom miniature mmWave radar attached to pixel phones for gesture classification. WiSee\cite{pu2013whole} was the first home-scale gesture recognition system for independent human location, followed by works such as \cite{zhang2021widar3, qian2018widar2, qian2017widar}. These systems either require customized hardware or can only identify a limited set of pre-defined gestures. Beyond gesture recognition, there has been some work on fine-grained human pose estimation using RF signals, such as WiFi, by analyzing the body's radio reflections. Adib and Zhao et al. did a series of work on human motion detection and pose estimation \cite{adib2013see, adib20143d, adib2015rf, zhao2018rf, zhao2018through} using costly software-defined radios (USRP). Jiang et al. \cite{jiang2020towards} then built a 3D human pose estimation system that uses commercial WiFi. ~\cite{xie2023mm3dface} employs mmWave for fine-grained face reconstruction. In summary, RF-based wireless perception requires custom hardware or huge bandwidth that is computationally expensive. However, our system uses existing speakers and microphones. And the sound speed is orders of magnitude slower than RF, which yields a great solution without wide bandwidth.

    
    
    

\subsection{Acoustic solution} 
Hand tracking systems using acoustic sensing mainly consist of gesture classification systems\cite{amaging2022, wang2020push, gupta2012soundwave, yang2023sequence, chen2022swipepass} and single-point localization systems\cite{nandakumar2016fingerio, wang2016device, mao2019rnn, liu2022acoustic}. 

For example, in SoundTrack\cite{gupta2012soundwave}, Doppler effect caused by the moving hand is used for classifying a limited set of course-grained pre-defined gestures. \cite{amaging2022} achieves 2D imaging of multi-parts, but they still only do gesture classification because of the limited imaging quality.
FingerIO\cite{nandakumar2016fingerio} is an OFDM-based finger tracking system that achieves an under-centimeter finger location accuracy and enables 2D finger drawing in the air using off-the-shelf mobile devices such as smartphones. Wang et al. \cite{wang2016device} propose a phase-based method known as Low Latency Acoustic Phase (LLAP) that allows for high-resolution localization of the fingertip. Mao et al. \cite{mao2019rnn} use a recurrent neural network (RNN) based method to localize the hand as a single point in a room-scale using MUSIC AoA algorithm. While these systems can continuously track a finger, they pick up the signal reflected from the nearest point of the hand/body, so they only measure the hand/body as a single point rather than recognizing the individual motions of multi-part joints. \cite{li2020fm} tracks multi-target from multiple hands but <=2 targets per hand.
In contrast, our system can track the multi-parts and detect all 21 keypoints to reconstruct a hand skeleton.

\section{Beyond-Voice System Overview}

Our system, Beyond-Voice, leverages the speaker and the microphone array in the commodity home assistant device to enable continuous and fine-grained 3D hand tracking. The key idea is to transform the device into an active sonar system and analyze the reflections to reconstruct a 3D hand continuously. Beyond-Voice represents the hand pose using the location of the 21 joins in the hand as shown in Fig.~\ref{fig:skeleton}

To achieve this, our system workflow consists of four main modules, as illustrated in Fig.~\ref{fig:workflow}. First, in the data collection module, we transform the device into an active sonar system transmitting inaudible FMCW, and a microphone array recording the reflections. Simultaneously, we use a depth camera, Leap Motion controller, to collect the ground truth hand position. Please note that Leap Motion is only required in training but not in final use. Secondly, the audio passes into the data preprocessing module, where we first clean the signal by removing the environment noise and other reflections that are not from the hand. We then derive a high-resolution range profile as a 2D feature map that represents the multiple joint positions. Next, the data from microphones merge into a multi-channel feature map and input to a deep learning model. This feature map along with the ground truth is used to train the model. At inference, the model can predict the 3D positions of 21 hand key points from the acoustic signal alone. In other words, the users do not need a camera on the device to use the hand tracking feature.

In the rest of the paper, We first describe our methods in detail. Then, we explain our experiment design and experiment results. Finally, we discuss potential use cases, limitations, and future work.

\section{System Design}
In this section, we describe in detail the key modules in the design of Beyond-Voice as shown in Fig.~\ref{fig:workflow}.

\subsection{Transform into an active sonar system}

The home assistant device commonly consists of speakers and a microphone array. Our system, without hardware modification, leverages these built-in acoustic sensors to transform the device into an active sonar. First, the speaker transmits a frequency-modulated continuous wave (FMCW) signal with a frequency of $17$--$20$KHz. It is inaudible and does no harm to humans as long as it is under 90db\cite{moyano2022possible}. Our ultrasound emits at 50db, a level that corresponds to a quiet conversation commonly encountered in daily life.

The microphone array record the reflected FMCW signal, which consists of multiple noise sources, including 1) any ambient noise of speech, music, etc., in the environment. 2) reflections from other objects in the environment. Therefore, we use the pipeline described as followings to separate hand's reflection from the noise.

\subsubsection{High-pass filter eliminates the audible noise}
\hfill

First, we apply a high pass filter to remove the signal below the frequency of 17k, which typically corresponds to background noise and speech noise under 300 Hz. This technique ensures that we have a preprocessed signal with a high signal-to-noise ratio (SNR), containing the desired FMCW reflections related to hand motions.

\subsubsection{Time-domain cross-correlation of FMCW yields high-resolution time-of-flight}
\hfill

Next, our system extracts the time-of-flight (ToF) of the signal traveling from TX to RX, i.e. speaker and microphone. Instead of using FFT as the traditional dechirping for FMCW Radar~\cite{instruments2020fundamentals}, we apply cross-correlation (xCORR) directly to the time domain which yields high resolution.

\begin{figure}[H]
    \centering
    \includegraphics[width=.85\linewidth]{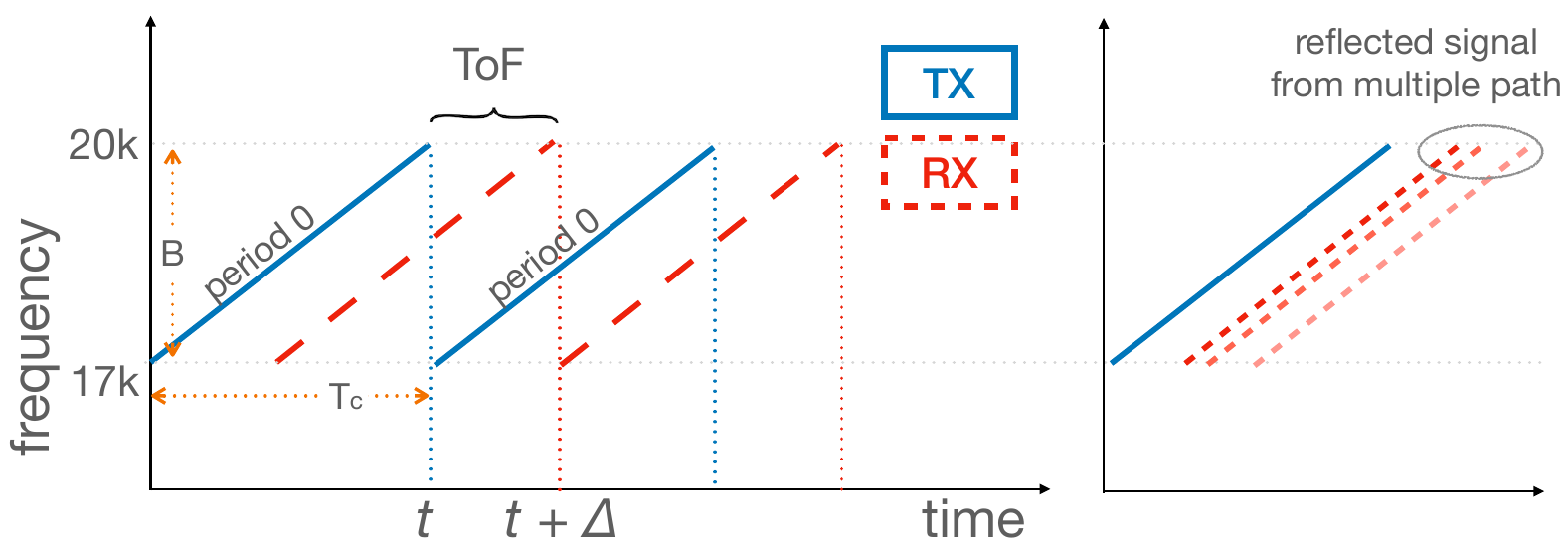}
    \caption{FMCW: each received signal is a time-delay version of the transmitted signal shifted by a different amount of time proportional to the distance.}
    \label{fig:fmcw}
\end{figure}

To elaborate, as illustrated in Fig.~\ref{fig:fmcw}, FMCW are basically repeated chirps that linearly increase in frequency. We choose a starting frequency of 17k, bandwidth ($B$) of 3k, and duration ($T_c$) of 512 samples (around 0.01s and 2m detection range) as one FMCW chirp. 
Let the blue line represent the transmitted signal (TX) at time $t$; the red line is the received signal (RX) that flies back at $t + \Delta$. This time delay $\Delta$ equals to $\frac{2d}{c}$, where $d$ is the length of one of the paths from hand to device and $c$ is the speed of sound in air. 

Next, the cross-correlation of TX and RX yields the time domain shift, which is indeed the ToF regarding a certain sample rate~\cite{wang2018c}. FMCW signal is properly correlated, which yields few side lobes, especially with a wide bandwidth of 3k. This is also an essential reason why we use FMCW cross-correlation as dechirping rather than the other traditional methods.
Finally, with this number of time shifts, sample rate, and speed of sound, we could get the absolute distance from objects to the device.

With this range measurement pipeline, the range resolution $\Delta d$ of our system is NOT the typical equation~\ref{equation:traditional-dechirp}~\cite{instruments2020fundamentals}. Instead, the range resolution of our approach is one order of magnitude smaller, calculated as equation~\ref{equation:our-dechirp}:
\begin{align}\label{equation:traditional-dechirp}
    \Delta d = \frac{c}{2B} = 343/2/3000 = 0.05717m = 57.17mm
\end{align}
\begin{align}\label{equation:our-dechirp}
    \Delta d = \frac{1}{fs} \times c \times \frac{1}{2}
    = \frac{343}{48000 \times 2}
    = 0.00357m = 3.57mm    
\end{align}
$c$ is the speed of sound in air. $B$ is the bandwidth of FMCW signal. $fs$ is the microphone sample rate.
In other words, theoretically, our system is able to detect minor motion not smaller than 3.57mm. This superior resolution essentially facilitates the downstream machine learning pipeline. In comparison, conventional FFT dechirping for FMCW, or methods based on Dopperler Shift or DoA, are insufficient in either resolution or the ability for 3D localization; FFT dechirping also usually requires specified mixer hardware.

In summary, the output of the cross-correlation is the range profile of the environment i.e. a sequence of reflected signal strengths at each distance where each peak corresponds to reflections from objects at that distance.

\subsubsection{Accelerate the computation of cross-correlation.}
\hfill

To accelerate the computation, we perform a frequency-domain cross-correlation which is faster than that in time domain~\cite{xCorrFD}.
Because shifting a signal in time domain is equivalently scaling it in frequency domain. First, a discrete Fourier transform (DFT) converts time domain signal $x$ to frequency domain $y$
\begin{align}
    y_k = \sum_{n=0}^{N-1} x_n e^{\frac{-j2\pi nk}{N}}
    = A_{k}e^{j\phi_k}
\end{align}
Then, a shift $\Delta$ in sequence $x$ maps to a scaling in $y$:
\begin{align*}
    x'_n &= x_{n - \Delta}, & y'_k &= e^{\frac{-j2\pi nk\Delta}{N}}y_k
\end{align*}
In essence, delaying $x$ for $\Delta$ samples in the time domain is equivalent to multiplying the DFT by $e^{\frac{-j2\pi nk\Delta}{N}}$. The window size is equal to the duration of FMCW chirp with no overlapping. 

\begin{figure*}
    \centering
    \includegraphics[width=0.8\linewidth]{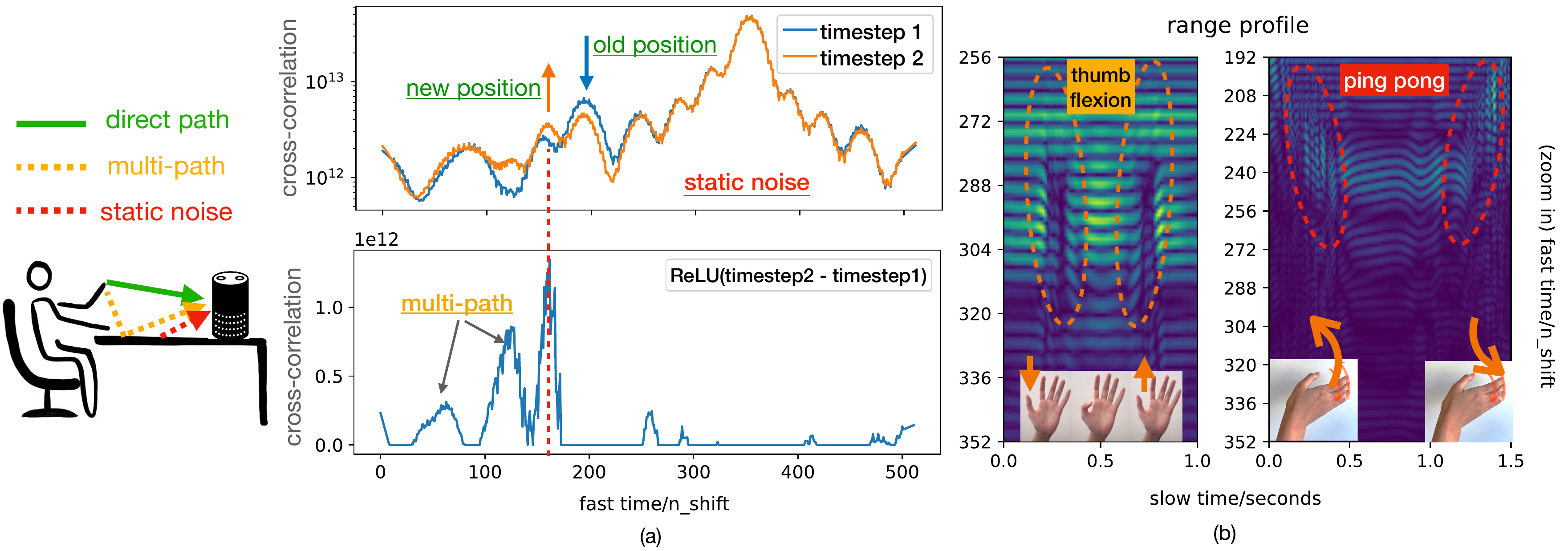}
    \caption{Signal processing: (a) Eliminate static noise by successive subtraction on cross-correlation windows. Each peak represents the strength of reflections $y$ from objects at distance $x$. (b) Align fast time along slow time to get a range profile. The exemplary spectrums show that continuous motions result in slashes + side lobes.}
    \label{fig:multipath}
\end{figure*}

\subsubsection{Successive subtraction removes the reflections from the static environment} 
\hfill

Upon having the range profile, the challenge is to remove the reflections from other static objects in the environment.

First, to eliminate the static noise, such as reflections from table surface, we subtract the cross-correlation output from consecutive time windows. As shown in Fig.~\ref{fig:multipath}(a), the static noise is consistent across time. By successive subtraction of the cross-correlation, the remaining peaks refer to the direct path of the moving object and its multipath. We use the Rectified Linear Unit (ReLU) to post-process the subtraction to simplify the information, i.e. we only keep the presence and ignore the corresponding absence. For example, when the fingers change position from timestep 1 to 2, the peaks of cross-correlation move, i.e., the reflection strength at different distances changes. The subtraction of two consecutive timesteps extracts this change so that the system can further localize the fingers. We assume the hand could not be absolutely static, since our range resolution is under-cm that can detect minor hand jitters.

Secondly, after removing the static noise, there is still a mixture of peaks. Some are the direct reflections from the finger; some are the multipath reflections reflected off both the moving fingers and the nearby static objects, for instance, the table surface. Because the multipath reflections always travel longer than the direct path, many existing works take advantage of this property by taking the first peak to only track the single closest finger point~\cite{nandakumar2016fingerio, wang2016device}. However, if aiming to estimate the multi-parts of hand, we cannot simply deselect the non-first peaks. Both direct paths and their multipath should be kept, which sum up to patterns that vary with the pose. Moreover, these complex patterns are no longer recognizable via rule-based algorithms, so we need deep learning in the following workflow.

\subsubsection{Hardware starting time cancellation}
\hfill

We program our speaker and microphones to start as synchronously as possible, which is essential to make sure that the shift in cross-correlation only comes from the time of flight. However, there is still a minor starting time delay each time we restart the speaker and microphones. What's worse, this delay is inconsistent; it varies across every restart. So, it is impossible to capture and eliminate it as a static noise factor.

\begin{figure}[H]
    \centering
    \includegraphics[width=0.9\linewidth]{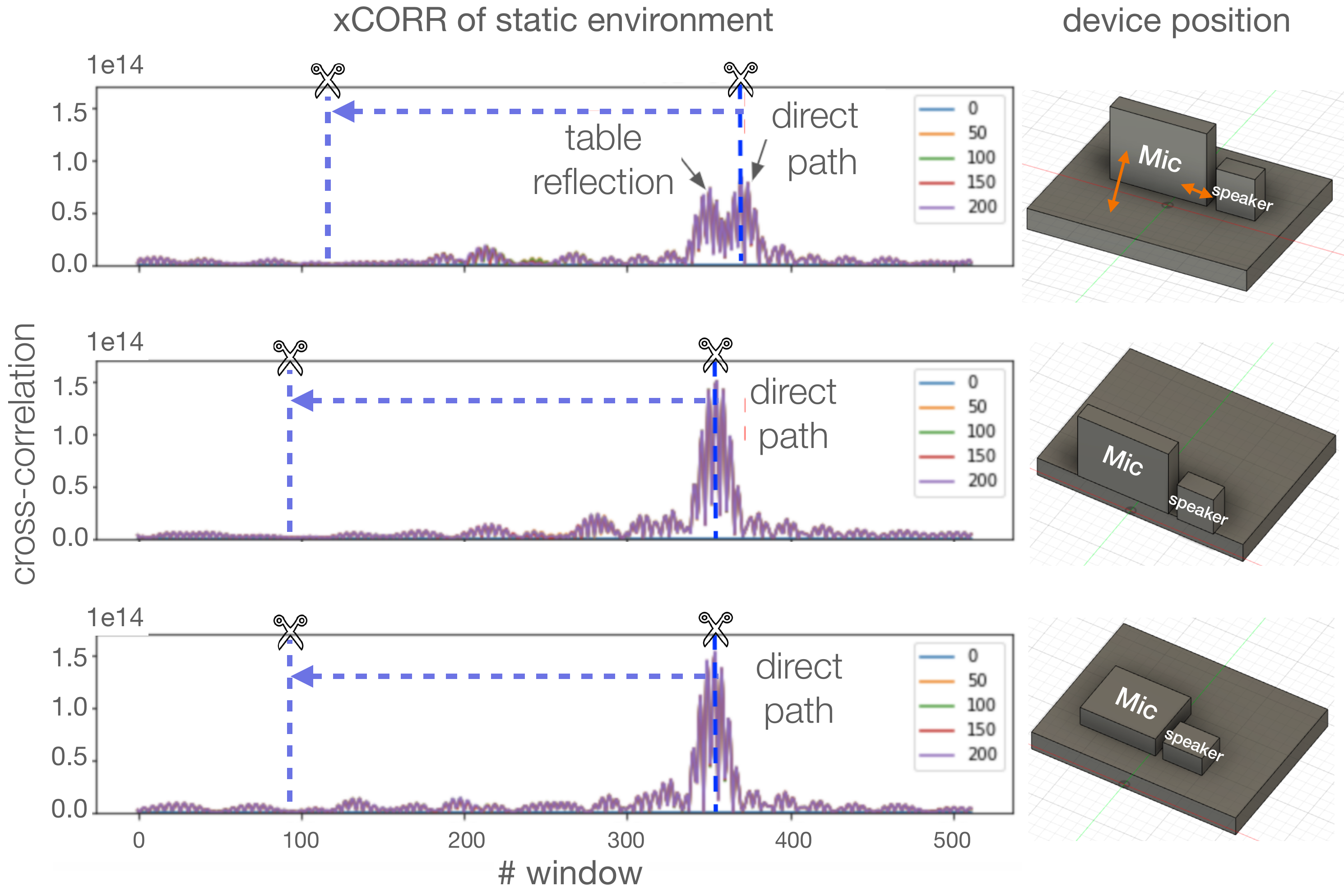}
    \caption{The starting time error caused by hardware restart is not fixed, but the direct path is usually the peak in different mic positions.}
    \label{fig:startingTime}
\end{figure}

To solve this problem, we find that the direct path can be used as an anchor to cancel the starting time error. This technique works because the direct path signal from the speaker to the microphone array usually has the strongest magnitude. For instance, Fig.~\ref{fig:startingTime} shows the cross-correlation (xCORR) output of sample windows at timestep 0, 50...200. (The lines of timesteps overlap together because there is no motion in the environment.) By altering the device position, the static environment slightly changes, and different peaks show up; as depicted, the direct path is always the highest and right-most peak in different microphone and speaker positions. Therefore, we can employ these two characteristics to distinguish it among the peaks.

In specific, to make the selection fault-tolerant, we select the top 10 peaks and then choose the right-most (nearest path). Then we cut off from the selected peak to the left for 256 steps, i.e., segment half of the original window, because it corresponds to \~1m range. This starting time cancellation algorithm is run only once after each time we restart the device. We apply this before applying the successive subtraction.

\subsection{Deep learning for range profile}
Multiple paths of reflections are entangled, which are delayed by a different amount of time proportional to the distance. Separating them using traditional closed-form algorithms is intractable. Even with deep learning, most previous work only localizes a single nearest point of the hand or up to 2 points per hand.

We aim to design a deep learning model to improve the detection granularity that can disentangle multi-part of the hand. The intuition behind this learning-based method is that our cross-correlation spectrums are feature maps for combinations of multiple reflections. If we carefully design the learning-based pipeline, it can use the range profile to estimate 21 joints in 3D.

\textbf{A classic model structure that captures spatial-temporal patterns:} In the Deep Learning module, we use a customized CNN + LSTM model architecture~\footnote{The system bottleneck is not the model as we tried some other model variants and the results remain similar.}. The high-level incentives are that CNN can model multi-channel spectrums at a single timestep, and LSTM can then combine a window of timesteps to capture the temporal pattern of consecutive motion. A high-level visualization of the model can be found in the deep learning module of Fig.~\ref{fig:workflow}.
The input size is (7, 256, 50), representing data from 7 microphone channels with a window size of 256 (half of the original cross-correlation window size because of the starting time cancellation), and a sampling of 50 windows for each data point, i.e., around 0.5 second, a duration sufficient to capture the finger motion. Next, the 50 windows are split into 10 folds (i.e. the input frame rate is roughly 19 frames per second, resembling the common frame rate for video streaming), and each is input into one CNN. Then we take the 10 outputs into an LSTM to capture the temporal pattern of consecutive hand positions. In other words, at every timestep, we look at the previous 0.5-second data to predict the current pose. In real-time inference, the input windows can overlap, so the output frame rate is not limited by the 0.5s input size but only related to the microphone sample rate.
In detail, firstly, each of the 10 folds in the shape of $(7, 256, 5)$ is input into the same CNN backbone separately. In the implementation code, we simply reshape the tensor by merging the batch size dimension and the fold dimension before passing it into the CNN; subsequently, we segregate the fold dimension again before feeding it into LSTM.
The CNN has two Conv2d layers, each followed by BatchNorm, ReLU, and MaxPooling layers. Secondly, an LSTM model is connected to capture temporal dependencies within the 10 continuous windows. The location information propagates across timesteps through the hidden state.
Finally, linear layers at the end frame the high-level feature vectors into a regression of 63 coordinates, which are the $(x, y, z)$ coordinates of 21 joints.

\textbf{Multi-channel input feature map from the preprocessed range profile: } To collect ground truth for training, we use a depth camera, Leap Motion, with its hand joints detection API\cite{leapmotion}. Its 3D coordinates of 21 joints act as the training label for the $7 \times 256 \times 50$ acoustic feature map. The ultimate output of the model is the 3D position of the 21 joints at a certain timestep, which can reconstruct the hand skeleton continuously.

The sample rate of the Leap Motion is dynamic, ranging from 90 to 110Hz. We timestamp each ground truth frame and align it with the high-sample-rate sound signals. It has 20 keypoints under the joints class, and another palm keypoint is added to keep consistent with the 21 keypoints ground truth that is widely adopted in hand tracking in the Computer Vision community. A visualization of the keypoint coordinate API is in Fig.~\ref{fig:skeleton}.

\textbf{Training strategies: data augmentation and pre-training} Overfitting is a common issue in pose estimation systems, including both human pose estimation and hand pose estimation, since the search space is large. Some works in the field of computer vision leverage intrinsic constraints of the human body kinetics to reduce the search space size. \cite{jiang2020towards} applies an initial skeleton and constrains the flexion angles to get the joint position. These methods require efforts of tuning and post-processing. Moreover, some methods lose the ability to detect the absolute position but only detect the relative position instead~\cite{zhang2020mediapipe}. 

Our system uses two strategies in training the deep learning model. One strategy is data augmentation which increases the training size with synthetic data. Another strategy is to pre-train the model with curriculum learning(CL). CL trains the model hierarchically from simple gesture sets to complex finger motions; otherwise, we observed that the model might converge at a static pose sometimes. Details are in the following two subsections.

\subsubsection{Data augmentation for overfitting reduction}~\\
\label{sec:dataAugmentation}
The performance of the deep learning system highly depends on the amount of training data collected. Since we predict the absolute 3D coordinates, the search space is huge, so it is hard to guarantee that the training data can cover every distance and every corner. To alleviate this challenge, we use data augmentation to generate pseudo data by slightly shifting the feature maps and the ground truth simultaneously. Since the spectrum represents the reflection distance, it is most sensitive to the change of y-axis. So, we shift the starting time cancellation cut-off towards the left (also slightly right). Each shift results in +/-3.5mm of ground truth y of all 21 joints and no change in the angle because the angles are relative positions depending only on the flexion of fingers. However, we cannot shift too much because the feature maps would have considerable changes besides just horizontal shift. To decide how much pseudo data we should generate, we test this factor with regard to the training efficiency and the improvement of performance obtained. By experiments detailed in \S\ref{sec: microbenchmarks}, we find that an augmentation factor of six works well with our system.

\subsubsection{Curriculum learning as pre-training}
\label{sec:curriculumlearning}
\hfill

When directly training on all gestures mixed, the task is so complex that the model might result in a high error and sometimes severely overfit, i.e., the output is a static hand. Therefore, we leverage curriculum learning (CL)\cite{bengio2009curriculum} to hierarchically bootstrap the training by allowing the model to start with learning simple tasks before complex tasks. With CL, once a simple task is trained, the model state is saved and then reloaded for training on a slightly more difficult task. To do this in our hand tracking tasks, we split mixed gestures into subsets by complexity and collect data for each subset separately. For example, we begin with moving a single finger for each of the five fingers, train on this one-finger data subset, save the model when training of this stage is completed, and reload the model to train on the two-finger data subset, which consists of moving every pair of neighboring fingers repeatedly. We repeat this process for increasingly complex motions involving consecutive triples, quadruples, and finally, quintuples of fingers. By increasing the number of fingers in iterations, the model successfully learns all gestures hierarchically with less chance of overfitting. In detail, it is intuitive that some poses are more difficult to track and error is different across poses. The gesture-wise mean square error (MSE) of individual to all finger motions are 276.67, 149.06, 359.83, 128.90, 41.67.
Note that we only apply CL to pre-training data. In the user study, participants do not collect data hierarchically but collect a mixed full set of gestures. We will explain it in more detail in \S\ref{sec:userstudy}.


\section{Implementation}
We evaluate the system with a user study of 11 participants and various benchmarks. This section introduces the setup, study design including pre-training, and evaluation metrics.

\subsection{Hardware and software Setup}

\begin{figure}[H]
    \centering
    \includegraphics[width=.7\linewidth]{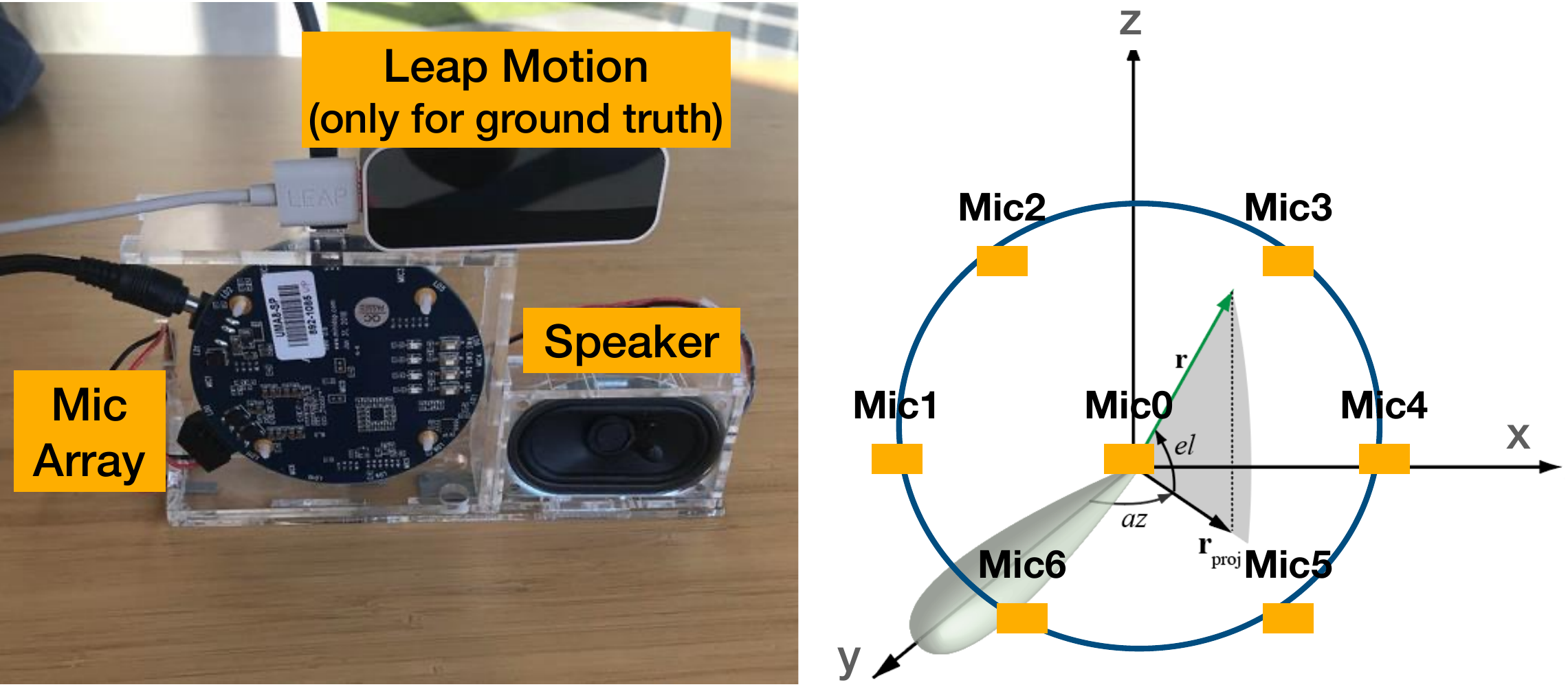}
    \caption{Hardware setup and the coordinate system.}\label{fig:device}
\end{figure}

\textbf{Hardware:} Our system consists of (1) a development microphone array board whose layout and sensitivity are the same as Amazon Echo 2 Home assistant~\cite{micboard}, (2) a speaker~\cite{digikeyspeaker} and (3) a Leap Motion infrared camera ~\cite{leapmotion} which is only for collecting ground truth in training. As shown in Fig.~\ref{fig:device}, a laser-cut acrylic holder puts them together. We use the development microphone board, similar to most previous work, because commercial home assistants do not provide API access to raw sensor data. Installing the raw firmware further eliminates the built-in DSP for optimal speech, which ensures the high frequency has a strong response. The board has similar sensitivity and layout to Amazon Echo Dot 2, which has a circularly arranged 7-channel UMA-8-SP USB microphone array. The speaker is a 4 Ohms General Purpose Speaker operating at 100Hz--20kHz range, soldered to the board. It plays inaudible FMCW ultrasound chirps, which then come back to the mic. The Leap Motion co-located with the microphone records the 3D locations of 21 hand joints as the ground truth. Note that Beyond-Voice does not require Leap Motion input in testing or in real use case. 

\textbf{Software:} The board is USB-connected to a MacBook running a Java program to record the raw audio. The transmitter and receiver lines operate on 24 bits per frame, PCM\_SIGNED encoding, at 48k sample rate, which are common settings in smart speakers and phones. We play an FMCW audio file generated using Audacity with 17k high-pass filter at the max roll-off(48db). The filter eliminates the audible burst sound at the intersection of two periods when a sudden frequency change happens. Leap Motion camera is connected to a Windows laptop running a Python program to record the ground truth hands and timestamps. A Flask RESTful API synchronizes with the Java audio program. Then we align the timestamps of camera frames and the starting time of audio frames. The data analysis is offline on an RTX v800 GPU server.

\subsection{User Study Design}
\label{sec:userstudy}

We conducted a user study with 11 participants (none of them are the authors) recruited from university campus. They are 5 males and 6 females of age $21$--$32$. The study was across three different environments - an office, a bedroom, and a small study room. The study was approved by the Institutional Review Board (IRB).

\begin{figure}[H]
    \centering
    \includegraphics[width=.75\linewidth]{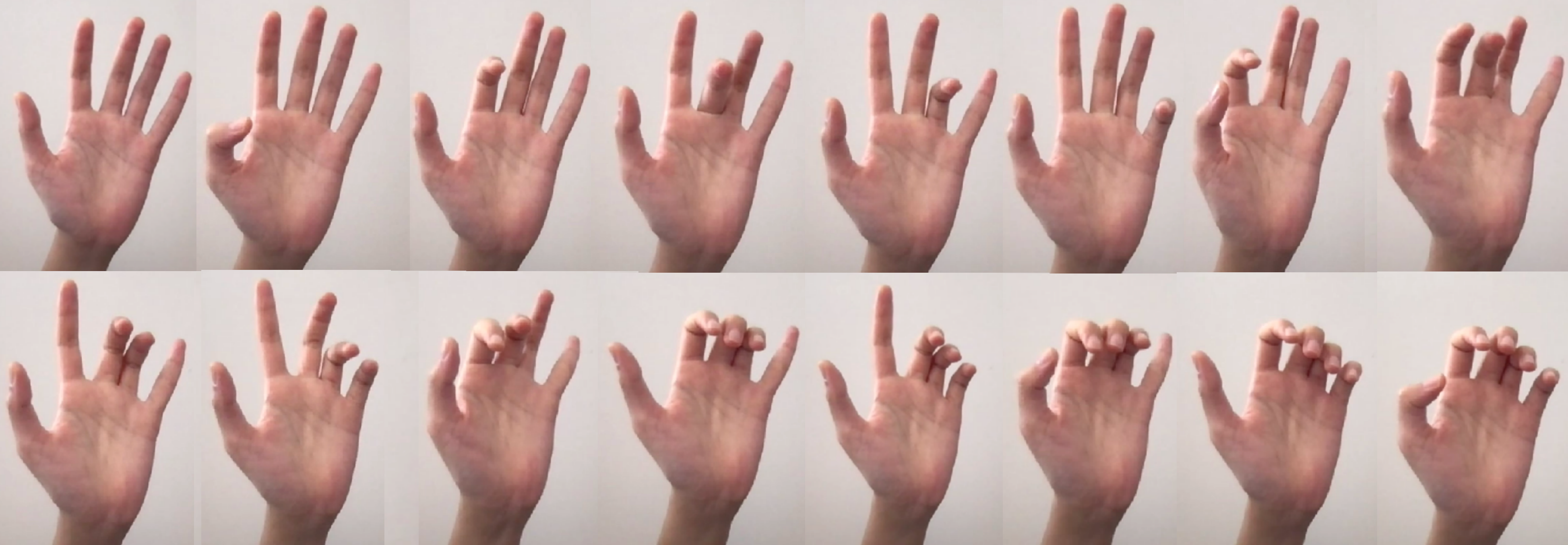}
    \caption{The demo poses in guiding video.}\label{fig:fingerMotion}
\end{figure}

\textbf{Data collection procedure. }
Participants are asked to sit at a desk and pose one hand in front of the device. The elbow and wrist could move freely in the air, and the device is on the desk. Each participant did two sessions. Each session lasts for 2 minutes.
In a session, they perform hand poses following the guiding video displayed on a monitor. It demonstrates a hand performing 15 gestures as Fig.~\ref{fig:fingerMotion}. Note that we do not do gesture classification, so the only purpose is to ensure the user can cover as many combinations of hand and finger motions as possible. Therefore, not strictly following the video is alright. In the meantime, we show a real-time visualization of Leap Motion hand tracking. Participants are told to place their hands roughly 20cm to 40cm away from the device for the best performance of Leap Motion. Talking is allowed since we will filter the recording in preprocessing.
Between two sessions, they are free to leave the desk and get re-seated. So the wrist position for the new session could be different from the previous. 
To test the system across environments, different users collected data in different environments. Most are in an office and some are in a study room or a bedroom. In total, we collected 64 minutes of data in 5 different days and 3 locations.

\textbf{Pre-training. }
As described in \S\ref{sec:curriculumlearning}, before the user study, we first pre-train hierarchically using curriculum learning. The pre-training data are from 2 additional subjects to minimize the efforts of user study participants. One subject was the author; The other was not involved in the system development; Neither is in the user study test. Each recorded ten data sessions, including five regular sessions and five simplified sessions. As aforementioned, a regular session contains a mixture of 15 gestures as Fig.~\ref{fig:fingerMotion}; A simplified session consists of flexing individual, consecutive doubles, triples, quadruples, or quintuples of fingers. We train the simplified sessions in the ascending order of the number of fingers, then the regular sessions, which largely outperforms that without CL.

\textbf{Evaluation metric}
We evaluate the average precision of estimating the absolute 3D positions of all 21 finger joints, between the estimated and ground truth. During training, to penalize errors in the large search space, we choose an $L2$ loss - mean square error (MSE) loss. While in our evaluation section, we report mean absolute error (MAE) as the main metric to facilitate error analysis across micro-benchmarkings. In perspective, we also compare with other modalities using mean-per-joint-position-error (MPJPE) in Euclidean distance. Furthermore, the outputs are visualized to substantiate the system performance. 

\section{Experiment Result}
\label{sec:experiment}
Following the user study design in \S\ref{sec:userstudy}, we evaluate our system performance and its generalizability across users, i.e. varies the amount of user's training data, including user-independent, user-adaptive, and user-dependent tests.

\begin{table}[H]
    \begin{tabular}{|c|c|c|c|}
        \hline
        & \textbf{mean} & \textbf{median} & \textbf{90th percentile} \\ \hline
        \textbf{user-independent} & 16.47 & 14.57 & 25.23 \\ \hline
        \textbf{user-adaptive} & 10.36 & 9.72 & 18.48 \\ \hline
        \textbf{user-dependent} & 12.49 & 10.33 & 21.41 \\ \hline
    \end{tabular}
    \vspace*{+2mm}
    \caption{Mean absolute error(mm).}
    \label{tab:mae}
\end{table}
\vspace*{-5mm}

\begin{figure}[H]
    \centering
    \begin{subfigure}[b]{.55\linewidth}
        \centering
        \includegraphics[width=\linewidth]{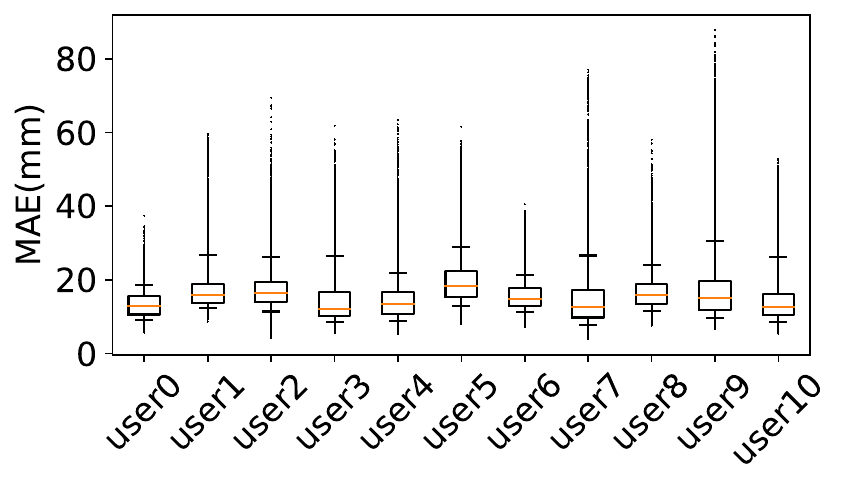}
    \end{subfigure}
    \begin{subfigure}[b]{.27\linewidth}
        \centering
        \includegraphics[width=\linewidth]{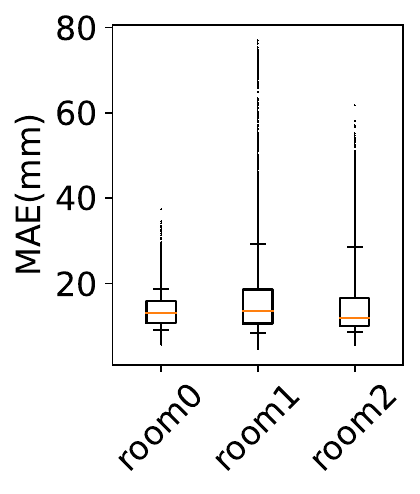}
    \end{subfigure}
    \caption{The error is independent on user and environment.}
    \label{fig:independenttest}    
\end{figure}

\begin{figure}[H]
    \centering
    \includegraphics[width=.95\linewidth]{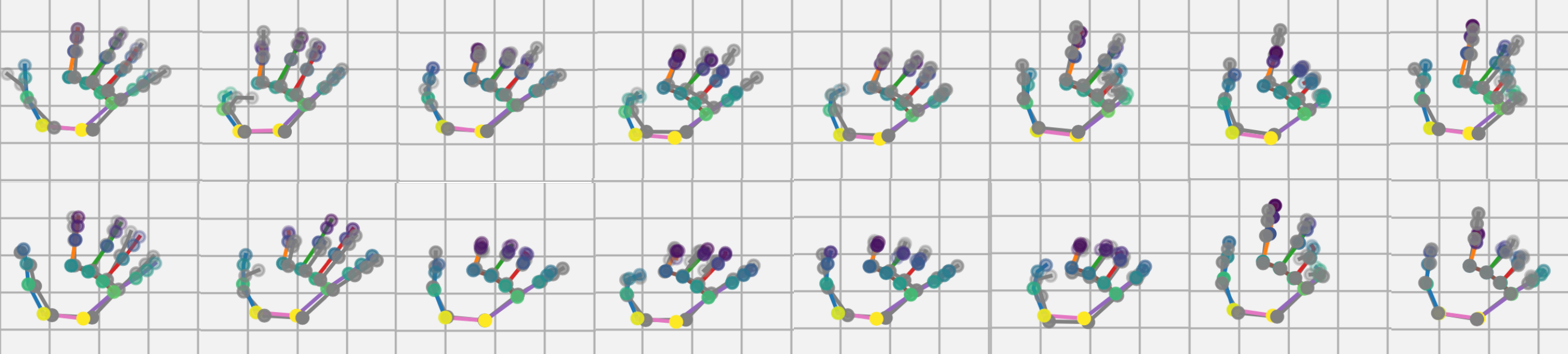}
    \caption{Sample results visualization: Grey skeleton denotes ground truth; cyan is our prediction (overlapped).}
    \label{fig:visFinger}
\end{figure}

\textbf{User-independent test:}
User-independent means the training set contains no data from the test user. In other words, we split the data into the training set and the testing set in a cross-validated leave-one-user-out manner. The pre-trained model also does not have any test user data there. This aims to demonstrate the system performance when there is no calibration/fine-tuning with user efforts. It simulates the actual use scenario when the off-the-shelf device might have no ground truth sensor to collect training data from a new user. The results of each test user are in Fig.~\ref{fig:independenttest}. The box plot depicts the median, 10th percentile, and 90th percentile of the error. All users' MAE yields average = 16.47mm, median = 14.57mm, and 90th percentile = 25.23mm. 
This user-independent test is the most challenging experiment in similar machine-learning-powered human sensing systems. Our result shows that the system performs well with no individual user training and is practical to develop and implement. Fig.~\ref{fig:visFinger} visualizes the result of sample hand poses, where the grey ground truth skeleton and cyan prediction skeleton mostly overlapped together. Moreover, post-processing techniques like smoothing can be employed to further minimize outlier errors.

\textbf{User\&environment-independent test:}
We further subgroup the user-independent folds by the data collection location to verify that the system works in unknown environments. We show that it could be trained and tested in different rooms since preprocessing steps remove the environmental noise. 
Recall that each user collected data in one of the three rooms: an open-space office, a bedroom, and a small study room. The average leave-one-room-out MAE is 15.73mm, as detailed in Fig.~\ref{fig:independenttest}, which is not higher than the benchmark of 16.47mm. So, the environment has no significant effects on system performance.
In the section of validation under interference, We further alter the nearby objects within the same environment using different materials including plastic and metal.

\textbf{User-adaptive test:}
Next, we evaluate the user-adaptive scenario where the system can collect training data from the test user, i.e., domain adaption or partially-leave-one-user-out. Although the system already works user-independently, there is a potential to collect some training data by the additional vision sensor in new form factors of home assistants. 
For each user, we added one of their session data to the pre-training data. The other session data was used as the testing set. As shown in Table~\ref{tab:mae}, the user-adaptive method could reduce the MAE to 10.36mm.

\textbf{User-dependent test:}
At last, we train and test with the same user's data and do not load the pre-trained model. The only purpose of this test is to benchmark the system with minimal effects of user or environment. 
To make it more valid, we need more than just two data sessions per user. So, 5 out of the 11 participants returned on another day to collect two additional sessions. Then one of the total four sessions is split out for testing, i.e., 4-fold cross-validation. As detailed in Table~\ref{tab:mae}, the MAE of the user-dependent test is close to the user-adaptive but slightly higher. The reason might be the smaller training set. Please note that we never split data intra-session. This ensures that device restart is always taken into account and simulates the actual use scenario.

\textbf{Comparison to literature:}
Since there is no work on 3D hand pose estimation from acoustic sensing yet, we put related literature of other modalities into perspective. Depth-camera-based 3D hand pose estimation systems achieve MPJPE typically ranging from 5.7mm to 20.8mm (on several public datasets including NYU, ICVL, MSRA) \cite{depthHandTrackingEvaluation}, depending on the specific dataset and methodology. \cite{ji2023construct} utilizes WiFi with an MPJPE of 22.1-27.1 mm on unseen users. Multi-view RGB images are mostly for root-relative hand tracking because their inherent depth ambiguity; recent work by Oculus VR~\cite{han2022umetrack, han2020megatrack} show promising results of absolute hand tracking with 11.2-13.6mm MPJPE on unseen users, while their online API currently only supports relative depth. Besides, the Human3.6M benchmark~\cite{zhu2023h3wb} has an MPJPE of 44.5-83.4mm for hands (aligned by wrist) at a whole-body motion scale. In comparison, our system has 21.7mm in MPJPE and 16.47mm in MAE, which shows a comparable performance and fidelity with a cheap audio-only modality. Fig.~\ref{fig:visFinger} shows the visualization of sample results.

\subsection{Micro Benchmarks Result}
\label{sec: microbenchmarks}

In this section, we break down the above results to benchmark the performance of the system for individual factors such as range, finger/bone, amount of training data, etc. The data for these analyses are from the above user study.

\begin{figure*}
    \centering
    \begin{subfigure}[t]{.22\linewidth}
        \centering
        \includegraphics[width=\linewidth]{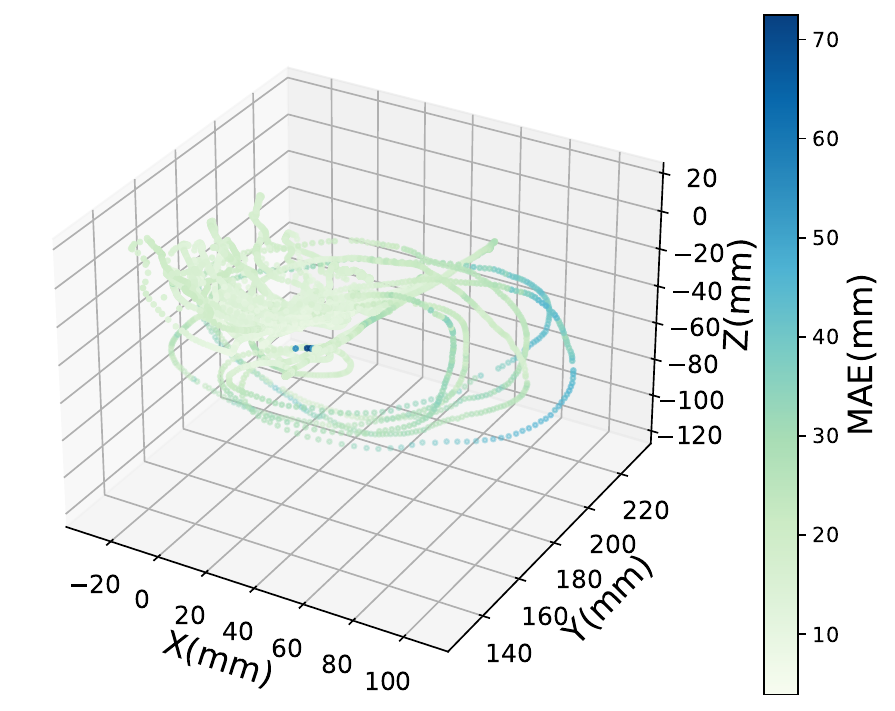}    
        \caption{Position of the wrist}
        \label{fig:distanceAcc}
    \end{subfigure}
    \begin{subfigure}[t]{0.18\linewidth}
        \centering
        \includegraphics[width=\textwidth]{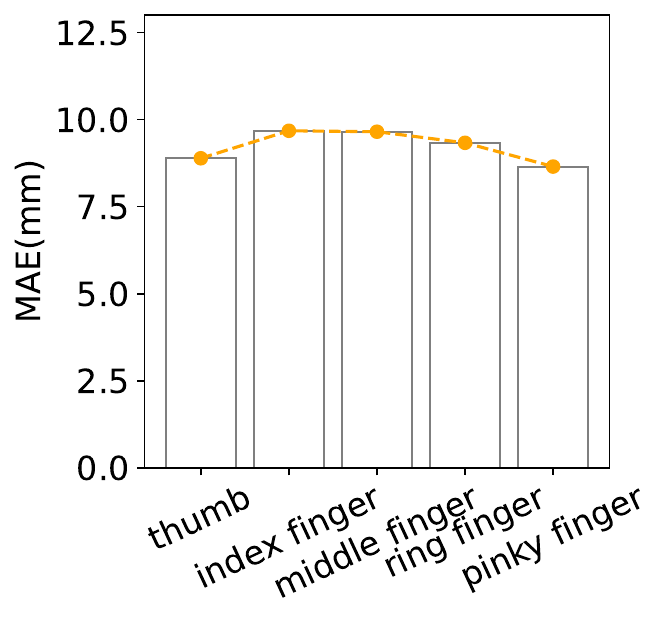}
        \caption{Finger-wise error}
        \label{fig:fingerwise}
    \end{subfigure}
    \begin{subfigure}[t]{0.185\linewidth}
        \centering
        \includegraphics[width=\textwidth]{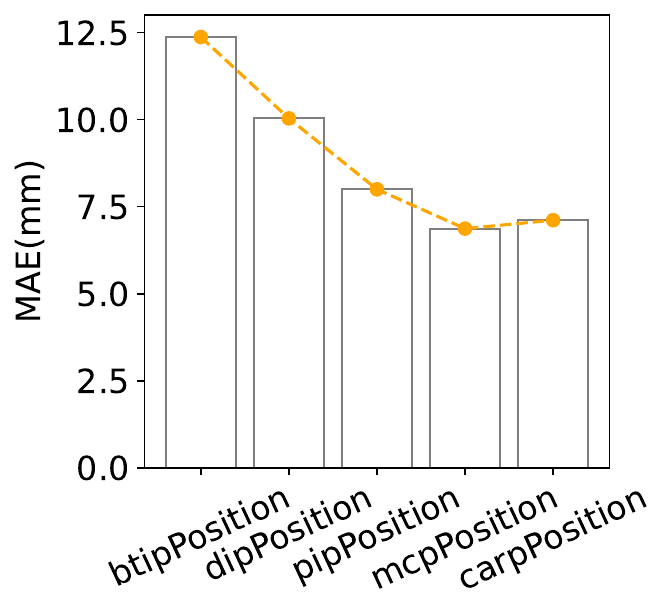}
        \caption{Bone-wise error}
        \label{fig:bonewise}
    \end{subfigure}
    \begin{subfigure}[t]{0.197\linewidth}
        \centering
        \includegraphics[width=\linewidth]{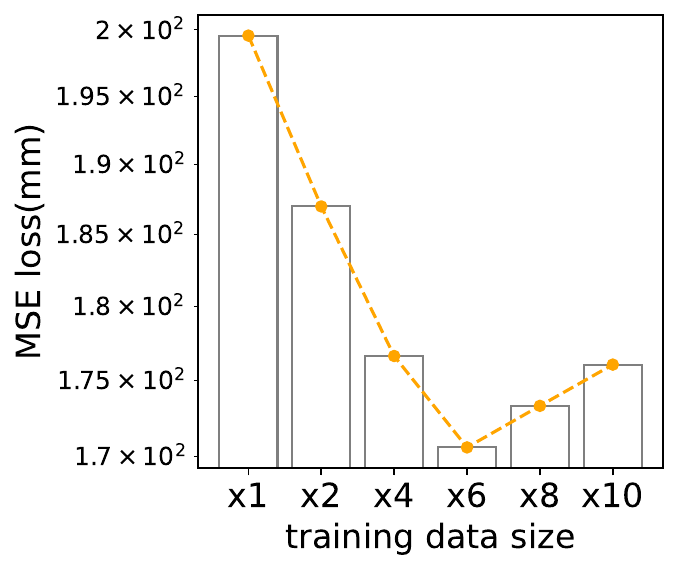}
        \caption{Data augmentation}
        \label{fig:generatedACC}
    \end{subfigure}
    \begin{subfigure}[t]{0.195\linewidth}
        \centering
        \includegraphics[width=\linewidth]{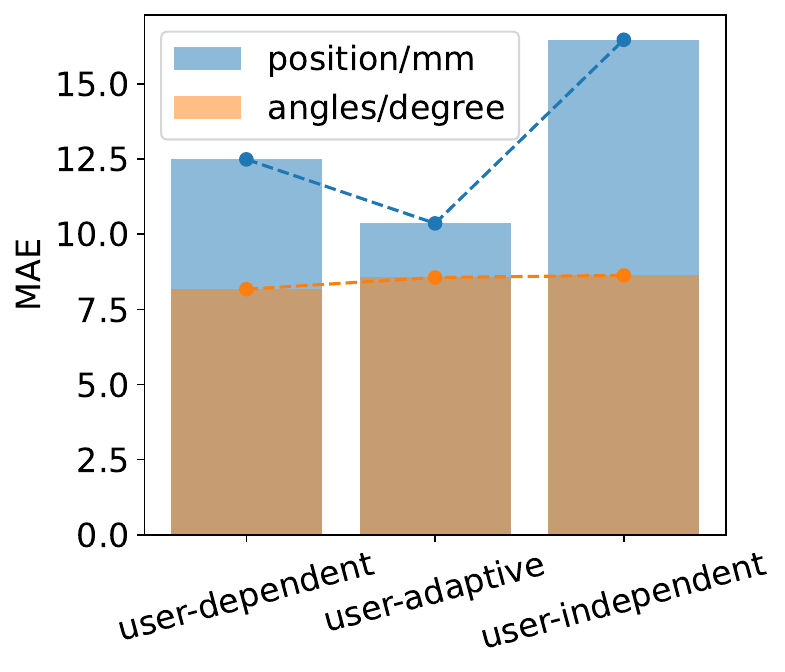}
        \caption{Finger flexion angles}
        \label{fig:angle}
    \end{subfigure}

    \begin{subfigure}[t]{0.17\linewidth}
        \centering
        \includegraphics[width=\linewidth]{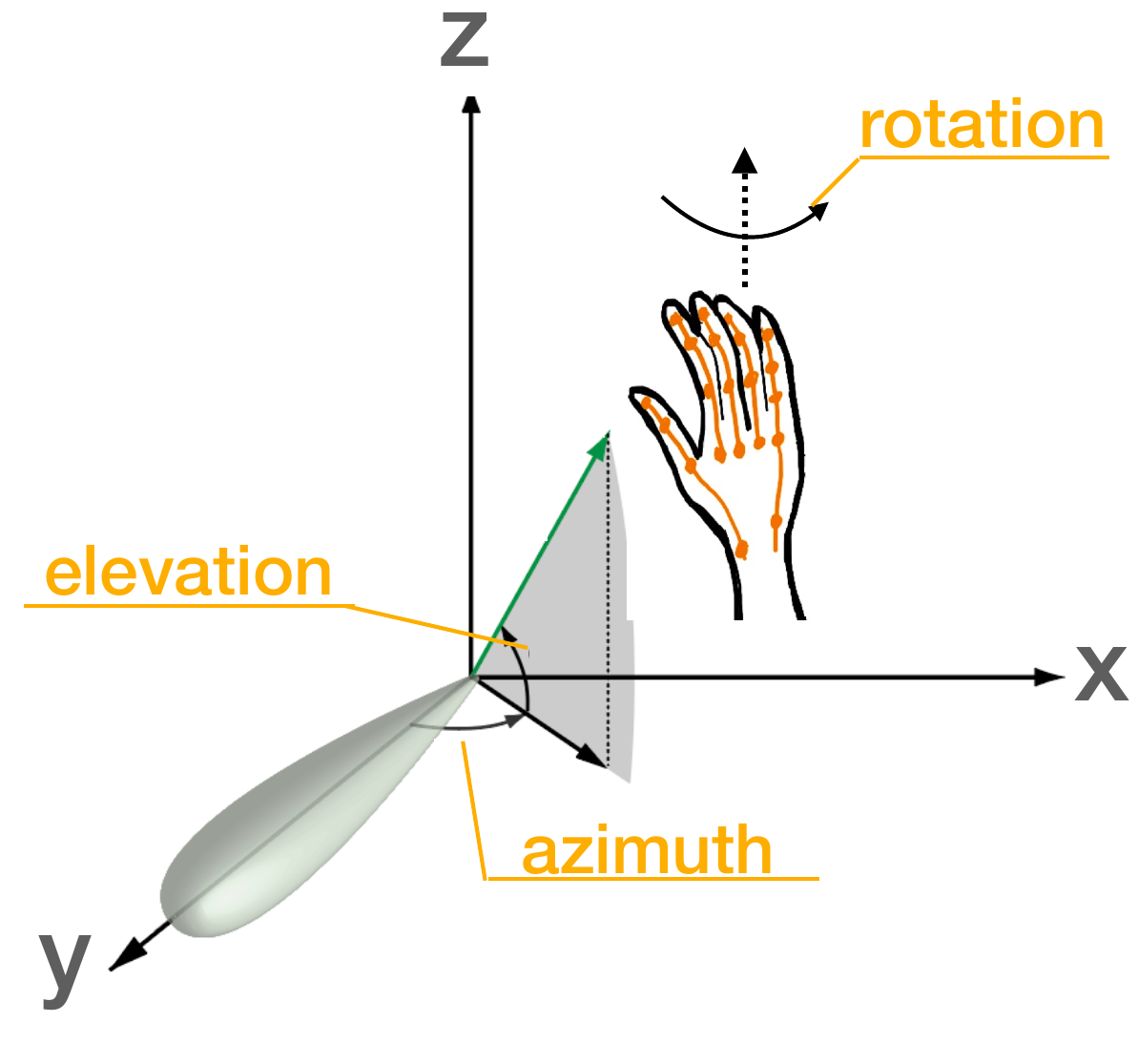}
        \caption{3 orientations of interest}
        \label{fig:orientationAxis}
    \end{subfigure}
    \begin{subfigure}[t]{0.2\linewidth}
        \centering
        \includegraphics[width=\linewidth]{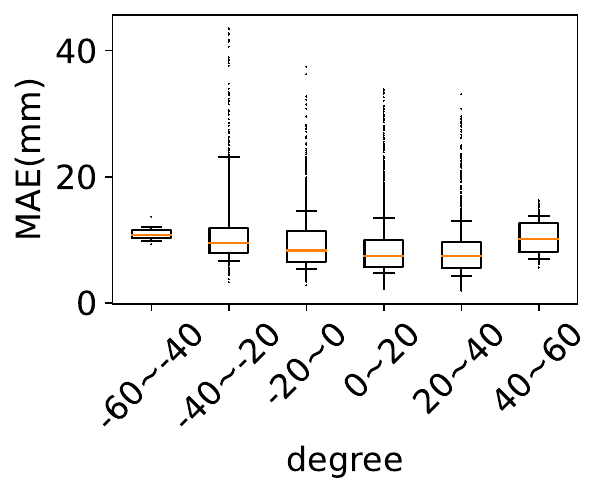}
        \caption{Palm rotation}
        \label{fig:palmRotation}
    \end{subfigure}
    \begin{subfigure}[t]{0.195\linewidth}
        \centering
        \includegraphics[width=\linewidth]{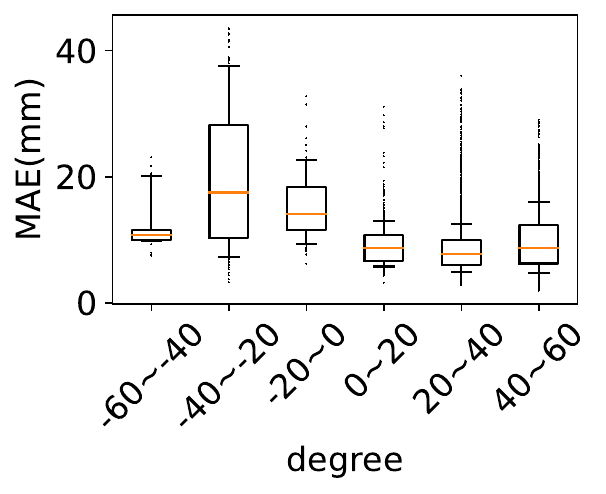}
        \caption{Azimuth}
        \label{fig:azimuth}
    \end{subfigure}
    \begin{subfigure}[t]{0.195\linewidth}
        \centering
        \includegraphics[width=\linewidth]{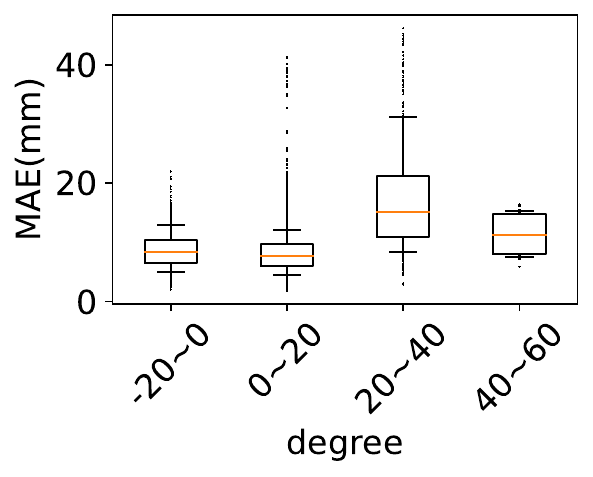}
        \caption{Elevation}
        \label{fig:elevation}
    \end{subfigure}
    \begin{subfigure}[t]{0.195\linewidth}
        \centering
        \includegraphics[width=\linewidth]{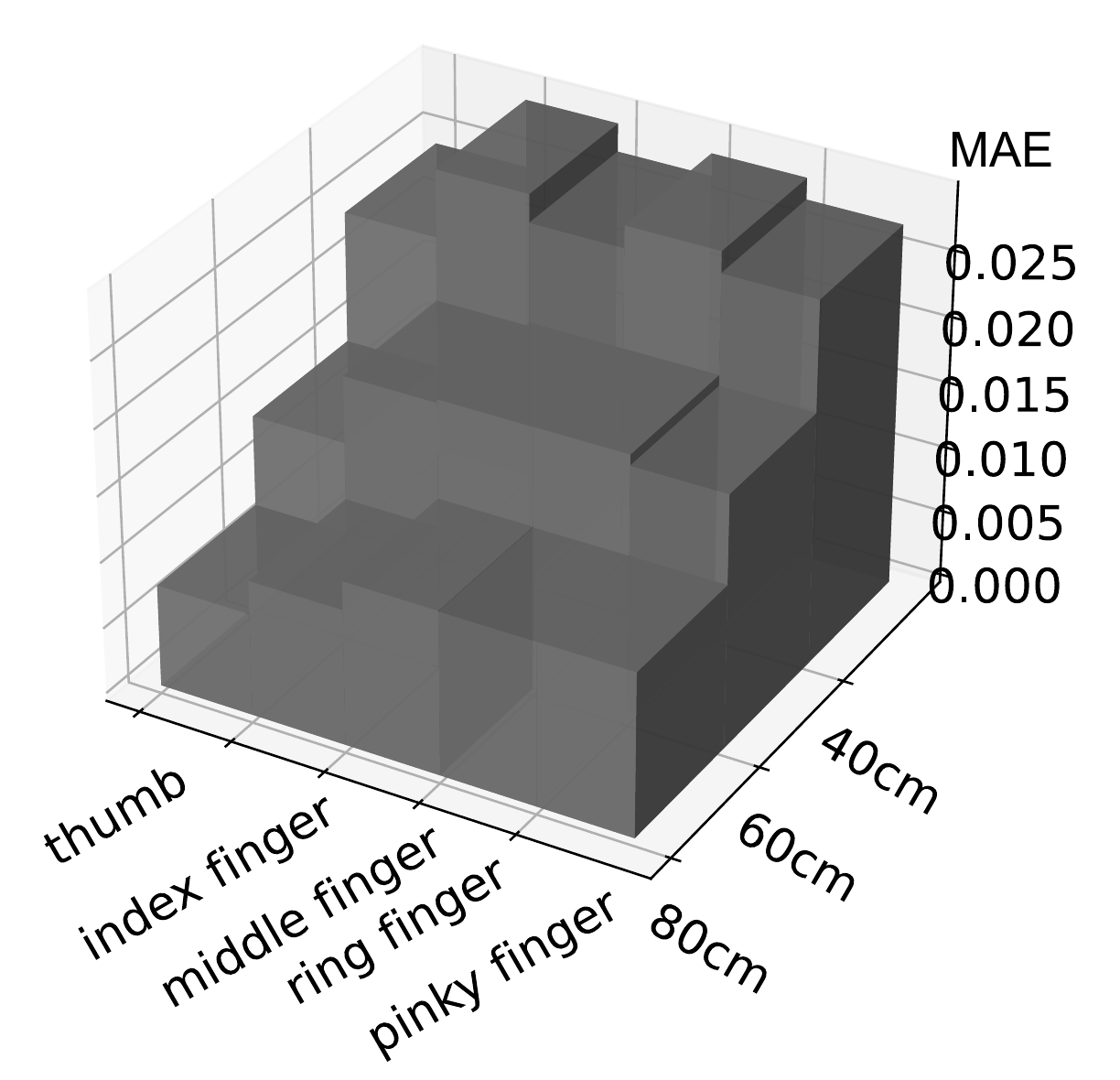}
        \caption{long range normalized to image size.}
         \label{fig:mediapipemae}
    \end{subfigure}
    
        \caption{Error analysis from different perspectives.}
        \label{fig:breakdown}
        \vspace*{-2mm}
\end{figure*}

\textbf{Effect of Range:} 
As illustrated in Fig.~\ref{fig:distanceAcc}, we analyze the error of the system with respect to the palm-to-device distance in 3D. The dots are the trace of the palm represented by the right carpal bone. Leap Motion is located at the original point in the coordinate system; the $x$-$y$ plane is parallel to the floor; the positive $z$-axis is facing down. The color of the dots is a visualization of the MAE error. 
From the data depicted in the figure, two observations arise. Firstly, clusters around $x=0$ consistently exhibit lighter shades regardless of the range in $z$ or $y$. This may be attributed to users predominantly positioning themselves around the $x$ center, resulting in more training data within this range, consequently leading to lower testing errors. Secondly, we note that although non-$x$-center outliers appear darker, their darkness diminishes notably when $y$ is within 150mm. Thus, we infer that the device distance $y$ holds greater significance than the positional parameter $x$.
Moreover, we acknowledge that data from multiple joints is impossible to be perfectly balanced per range. Therefore, to isolate the effect of range regardless of data distribution, we adopt hardware with longer range capabilities in section~\ref{sec:longer-range}, allowing us to coarsely group the data by solely the controlled wrist position.

\textbf{Effect of orientation:}
Per the effect of orientations, we analyze three orientations of interest including palm rotation, azimuth, and elevation as illustrated in Fig~\ref{fig:orientationAxis}. 
We group the errors by angles ranging from -60 to 60 degrees, which is the sensitive range of our hardware (except >-20 for elevation because it is below table). Although some groups of large angles show high error in Fig.~\ref{fig:azimuth} and ~\ref{fig:elevation}, there is no significant correlation between average MAE and orientations. Therefore, we could assume that, if within the hardware sensitivity and with enough data, orientation does not have a significant impact on system performance. This proves the advantage of wireless sensing compared with cameras: camera’s performance usually decays in extreme rotations that have high self-occlusion, while wireless signal can capture the non-line-of-sight by penetrating soft materials.

\textbf{Finger/bone-wise accuracy:} 
In this experiment, we break down the error of user-dependent results to analyze the system performance for individual fingers and joints. Fig.~\ref{fig:fingerwise} depicts that the fingers in the middle(index finger, middle finger, and ring finger) have a bit higher error. The reason could be that they will have slight movement when we intend to move their neighbor fingers, which adds entropy to the overall motions of the middle fingers.
Fig.~\ref{fig:bonewise} shows that the closer to the fingertip the bone is, the higher the error is. Because the fingertip's rotation radius is larger, namely the movement distance is more significant.

\textbf{Effect of increasing the size of training data with data augmentation:}
As detailed in \S\ref{sec:dataAugmentation}, data augmentation is an efficient way of adding generated training data. To evaluate the effectiveness, we compare the MSE loss under different scales of data augmentation. Fig.~\ref{fig:generatedACC} depicts that the error decreases as more data is generated until it reaches five times the original amount. Then it experiences a slight rebound when increasing from x6 to x10. This rebound may indicate overfitting from the perspective of machine learning; While regarding signal processing, the rebound could result from over-shifting of the spectrum for augmentation. To elaborate, due to the multi-path effect, large movement will induce non-negligible changes in the range profile other than just shifting.
Therefore, we choose x6 as the scaler of data augmentation to gain the best performance. In other words, generated data are within 1cm of the original data points.

\textbf{Flexion angle estimation:}
Seeing that Leap Motion can also record finger flexion angles, we also tested predicting flexion angles instead of the joints position. 
Angles captured are Leap. Vector, which is the z-basis of each bone, including distal, intermediate, proximal, phalanges and metacarpals. We calculate the flexion angle per bone by taking the angle between the current bone and its ancestor towards the wrist. For the root bone, the ancestor is the unit vector of the negative z-axis, which is the norm of the table. Since the thumb has no metacarpals, there are 19 angles in total. The deep learning model is the same except for the size of the last linear layers.
We run user-dependent, user-adaptive, and user-independent tests using angles the same as using positions. In Fig.~\ref{fig:angle}, the error stays stable across users. It aligns with our intuition that angles are less prone to the change of hand size.
Compared with 63 coordinates, 19 angles reduce the search space, making training more accessible. However, angles are relative position which is a subset of absolute position. We prefer the latter because it enables more rigorous hand tracking such as gaming and placing objects in VR/AR.

\textbf{Speed of motion:}
In theory, the Doppler effect does not affect our cross-correlation-based ranging accuracy, unlike traditional FMCW dechirping method~\cite{instruments2020fundamentals}. Because Doppler effect causes additional frequency shift, but our time shift is always proportional to the instant distance regardless of speed. Besides, the potential windowing distortion is trivial as long as the speed of the finger is less than the speed of sound. We verify this by calculating the Pearson correlation coefficient between speed and error of our results, which is -0.0027, indicating no correlation.

\subsection{Validation under Different Interference}
In this section, we validate the robustness of Beyond-Voice in the presence of different environmental noises, including audible noise, other motions in the environment, and nearby objects and metal. 

\begin{figure}[H]
    \centering
    \includegraphics[width=.95\linewidth]{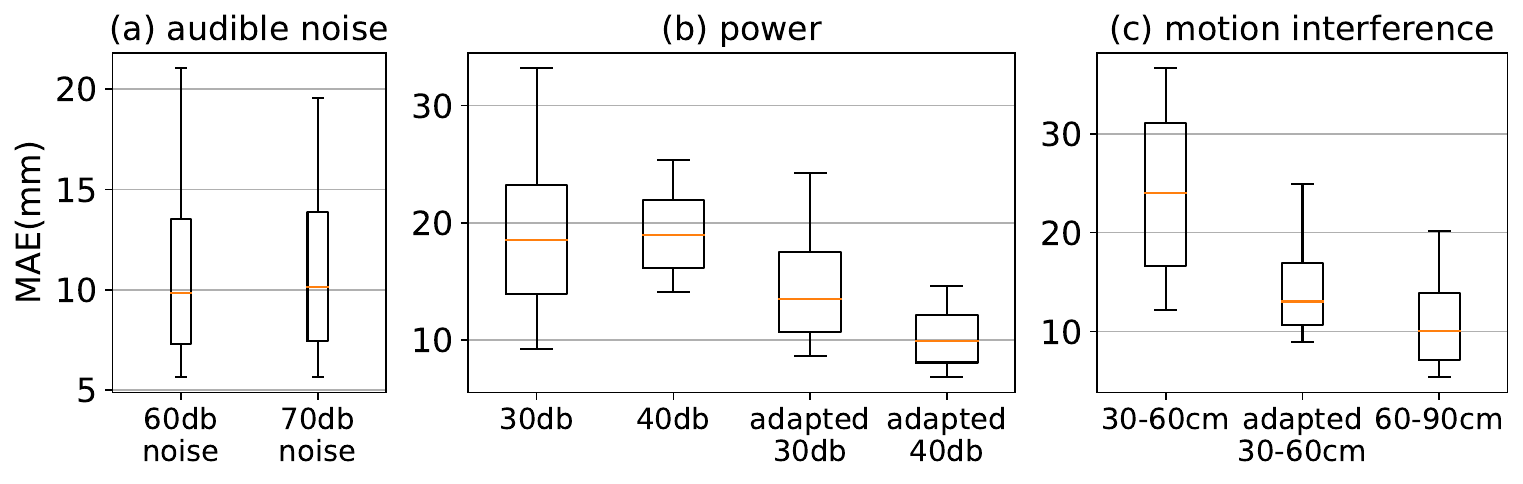}
    \caption{(a) The audible noise does not affect the system performance. (b) The accuracy drops when ultrasound volume is <50db. (c) Nearby motion interferes the accuracy. (b, c) But adaptive training helps.}
    \label{fig:validations}
\end{figure}

The hypothesis is that the system performance should remain unaffected by audible ambient noise; the detection accuracy could drop with the presence of another moving object or metal nearby, or when the ultrasound volume is low. 
To evaluate each of these conditions, we conducted a new set of experiments with three participants from the user study. Each user performed two additional sessions for six scenarios. Training data are their user study data along with pre-training data, and the new data are for testing or domain adoption. So, we use the user-adaptive result in Table~\ref{tab:mae} as the baseline. Results are shown in box plots in Fig.~\ref{fig:validations} with the median, 10th percentile, and 90th percentile of MAE.

\textbf{Concurrent audible noise/playback}
Our system allows concurrent audible noise from the same device or the environment. 
First, our high pass filter in preprocessing removes the audible ambient noise below 17kHz without polluting the ultrasonic spectrum when the sample rate is 48kHz. To validate this, we play a Youtube video of Ellen Show, i.e., conversational speech, in the background. As shown in Fig.~\ref{fig:validations}(a), the audible noise at 60db and 70db yields an average MAE of 11.76mm and 11.99mm. (60db is the volume of daily conversational speech. 70db is as loud as a vacuum cleaner.) Besides, in the aforementioned user study, we also allow the user to talk during recording.
Secondly, we avoid using high-power ultrasound, unlike many other works that typically rely on this to extend detection range. Therefore, concurrent music playback in the same speaker won't be distorted because our $\leq$50db ultrasound can hardly crowd the mixer \cite{li2022experience}, thereby allowing concurrent playback of up to \~66db.

\textbf{Power of FMCW signal:}
The volume of the FMCW ultrasound is 50db in our default settings, which is the volume of the average home. In Fig.~\ref{fig:validations}(b), we turn down the ultrasound to 40db and 30db, thus the errors increase to 19.59mm and 20.23mm. Therefore, we use domain adaption to fine-tune the model. In detail, by adding one session of 40db in training data, the error of 40db returns to 10.58mm, which is at the same level as the baseline. Moreover, the error of adapted-30db test drops to 15.75mm. In conclusion, if the power of ultrasound is 30db--50db, the system performs well with adaption in training. 

\textbf{Motion interference:}
Although we can remove the reflections from the static furniture, our system is susceptible to moving objects other than hands. In theory, the sound rapidly attenuates when traveling in the air, so motions beyond 1 meter away should not affect the system. Thereby, we ask another person to move around at 30cm\~60cm and 60cm\~90cm away. Fig.~\ref{fig:validations}(c) shows that the MAE of having interference beyond 60cm is 11.94mm, which is slightly higher than the benchmark. As for <60cm, the MAE increases to 24.02mm. By adding data with <60cm motion interference in training, the adapted model results in an MAE of 18.25mm (median \~13mm), which is better than the non-adapted. 

\textbf{Nearby occlusion using plastic and metal objects}:
Despite testing the cross-environment ability by altering the room, we also alter the nearby objects as line-of-sight occlusion: with no nearby objects, with a plastic cup nearby, and with a metal cup nearby per session separately. The result shows that plastic cup induces an MAE of 12.94mm which is close to the scenario with no nearby inference. Conversely, having a metal cup nearby leads to an MAE of 19.56mm. This outcome aligns with the fact that metal, being one of the most reflective materials, poses a common challenge in wireless sensing systems.

\subsection{Longer-range test}
\label{sec:longer-range}
During the user study, subjects were limited to a suggested range of 40cm due to Leap Motion's limited detection range as ground truth. However, if we simply move the palm 1 meter away from the device, we could still observe a clear change in the range profile. Hence, to verify the real detection range of our acoustic sensing, we replaced the Leap motion camera with an RGB camera as the ground truth and tested for a longer range. However, the RGB camera could only provide 2.5D position, which is extracted by Google MediaPipe hand tracking\cite{mediapipe} model API.

Compared with Leap motion's 3D coordinates, the 2.5D is defective 3D relative to the image settings and wrist position. Specifically, \textit{x} and \textit{y} are normalized to [0, 1] by the image width and height. \textit{z} is the depth with the wrist being the origin scaled roughly by \textit{x}. We collected 6 sessions from 1 user who was not involved in the system implementation, to test the user-dependent performance. And we do not split the data for training and testing from the same session. A longer range up to \~80cm was tested.

Since \textit{x} and \textit{y} are normalized to [0, 1], their MAE is measured proportional to image width and height; the camera resolution is 1920 x 1080. The average MAE at \~40cm, \~60cm, and \~80cm are 0.028, 0.019, and 0.011, i.e. around 2.8\%, 1.9\%, and 1.1\% of image size.
The finger-wise MAE in Fig.~\ref{fig:mediapipemae} indicates a decrease in error with distance, attributed to the hand appearing smaller in the image as it moves farther from the camera, and the measurement being proportional to image size. To remove this effect, we normalize the error by palm size and get errors of 0.1257, 0.1255, and 0.0991, i.e. \~9-12\% of palm size.

These results demonstrate the feasibility of our system at longer ranges. It also shows the potential for training using commonly available RGB cameras. However, it is important to acknowledge that RGB camera is not an ideal ground truth because 2.5D is not accurately synchronous to the acoustic signal change. For instance, when the palm move towards the device, the absolute distance changes while the \textit{z} in 2.5D relative to the wrist remains unchanged. So, ideally, to further extend the range, the costly room-scale motion capture systems might be better ground truth.

One meter is a range comparable to other acoustic sensing systems using COTS devices; a customized loudspeaker~\cite{nandakumar2017covertband} might further increase the range. We understand it is less than the range of voice interaction, but our goal is to compensate the VUI for accessibility and usability. Compared to existing gesture input systems which require users to touch the device with simple gestures like tap and swipe, extending the interaction range from 0m to 1m with a much richer expressiveness of tracking will significantly improve the interaction experience at zero cost on hardware.

\subsection{Enabling Demo Applications}

\vspace*{-3mm}
\begin{figure}[H]
    \centering
    \includegraphics[width=.9\linewidth]{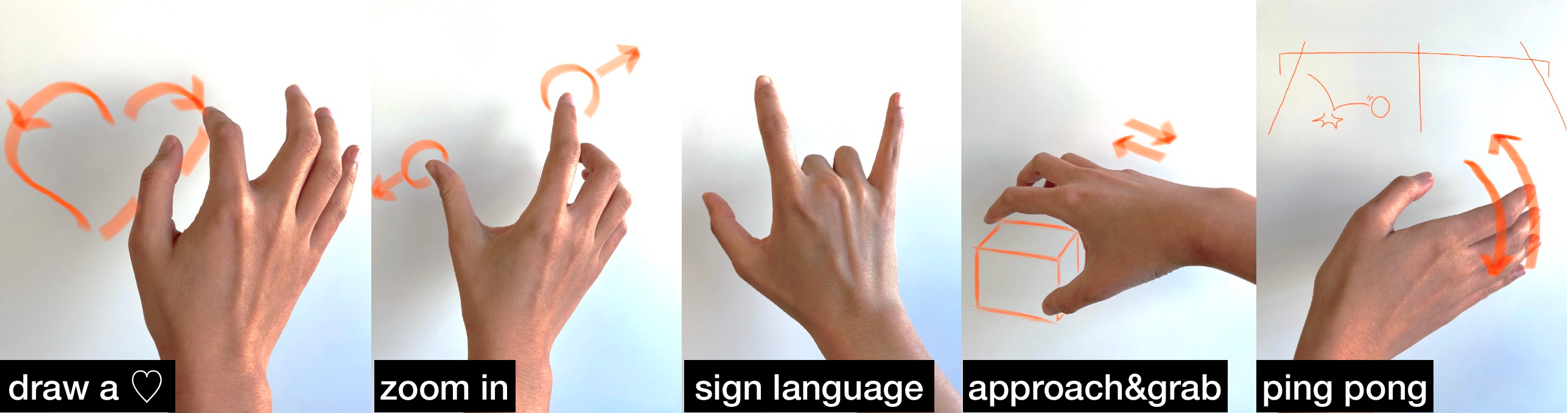}
    \caption{Demos of continuous gesture tracking.}
    \label{fig:demos}
\end{figure}

To better understand the usability of Beyond-Voice, we test certain intuitive applications that need continuous and absolute-range hand tracking, such as i) \textit{drawing in the air} as an alternative input space ii) \textit{zoom in} as a hand gesture to increase or decrease volume iii) \textit{sign language} as an interface for people with speech disorders iv) \textit{grabbing and placing objects at a specific position} in VR/AR application v) \textit{playing ping pong} for gaming apps.
Fig. ~\ref{fig:demos} illustrates the five motions. Their MAEs are 11.70mm, 15.54mm, 13.48mm, 15.54mm, and 19.99mm respectively. \href{https://drive.google.com/file/d/1CfFvV-cCkutmsynIu1LmDuNROpkfAP9L/view?usp=sharing}{And a supplementary video}~\cite{demoVideo} visualizes our results at a x10 subsampled frame rate. Note that these demo gestures are absent in the pre-training dataset.

The key advantage of our continuous tracking is that it is not bound to pre-defined gestures. Developers can directly use it as a versatile API. Secondly, continuous fine-grained tracking is necessary for scenarios like dragging, drawing in the air, zoom-in, etc. Besides, gesture applications, like sign language recognition, can work on top of continuous tracking. Also, as our system detects the absolute coordinates, it provides more information than existing gesture recognition systems; for instance, playing ping pong game with the smart speaker as an on-table sensor/controller.

\textbf{Gesture recognition as system activation: }
We envision that our system will have an activation pose in real-world deployment (equivalently “Hi Siri” for iPhone). To prove that, we take the sign language of ‘love’ in Fig.~\ref{fig:demos} as an activation pose and test it on existing data. By calculating the similarity between each predicted skeleton and the predefined ‘love’ skeleton, we can enable activation pose detection using a simple similarity threshold, with a 99.9\% accuracy. In detail, we achieve this by aligning the orientation and size of the palm by rotation matrix and normalization, then defining similarity as negative MAE. The average MAE of ‘love’ is 0.002 (99.9th-percentile=0.004), while all other poses yield 0.111(0.1th-percentile=0.085). The significant gap between 0.085 and 0.004 ensures that a threshold is reliable to distinguish the presence of activation pose.
This result demonstrates the feasibility of incorporating our continuous tracking as part of a gesture recognition system. As for more gestures, developers can further employ end-to-end methods as well.
\section{Discussion and Future Work}
\label{sec:discussion}
Our work indicates a pathway for enabling fine-grained 3D hand tracking by leveraging acoustic sensors. In this section, we discuss the limitations and potential future work.

\textbf{Implementing Beyond-Voice on off-the-shelf home assistants:} In general, our system is a software solution and requires no hardware modification, so a wide variety of commodity home assistants could potentially adopt this framework as long as they train with their own devices. 
Their settings, such as the number of acoustic sensors and their arrangement, might be diverse which could potentially affect the performance. However, we also note that the current trend shows an increase in the number of microphones (Google Sonos has six microphones, three more than the old Google Home) that will only improve the performance of our system. Besides, different frequency responses and case wrap might affect the signal power. However, Fig.~\ref{fig:validations}(b) validate that it works well when volume decreases by 20db(around x$2^6$ less in power).

\textbf{Tracking both hands:}
At present, we only train and test our system on the right hand. However, we expect that the left hand would perform similarly if sufficient training data from left-handed individuals is available. We believe that home assistant manufacturers have the resources to conduct extensive training under various conditions. Additionally, tracking multiple hands is a challenge because of more occlusion, that we leave as future work.

\textbf{Real-time implementation of Beyond-Voice:}
The model parameter size is 3.53MB. The estimated forward/backward pass size is 18.46MB. So, it fits in commercial smart speakers whose memory is usually at least 500MB. Its on-device inference time is 10.4-20.8ms, profiling on the Echo Dot 2 featuring ARM Cortex-A53. And that is 0.4ms on a GPU server. Hence, deploying this small model in real-time should be achievable for on-device computing. Moreover, advanced model compression and hardware acceleration may further improve on-device computing and minimize uploading user data to the cloud.

\textbf{Potential VR Applications:}
Virtual reality (VR) is also an area that can benefit from a non-contact near-field hand tracking system. The latest Oculus headset released free hand tracking without controllers, allowing immersive interaction. Most of these headsets utilize visual-based sensing. However, vision is prone to (1) light conditions, (2) line-of-sight/occlusion, where the hand is a highly self-occluded case, and (3) privacy concerns. Thereby, acoustic sensing provides an alternative or supplementary solution.

\section{Conclusion}
In this paper, we present Beyond-Voice, a novel acoustic-sensing system, which enables continuous 3D hand pose tracking on home assistants by leveraging the speaker and the microphones on device. A user study with 11 participants showed that Beyond-Voice could reconstruct the 3D positions of 21 finger joints with an MAE of 16.47mm without using any individual training data from the user. 


\begin{acks}
We sincerely thank the anonymous shepherd and reviewers for their constructive suggestions. This research was partially supported by National Science Foundation Grant No. 2239569.
\end{acks}

\bibliographystyle{ACM-Reference-Format}
\bibliography{main}


\end{document}